\documentclass[aps,prl,reprint,superscriptaddress,bibnotes,eprint]{revtex4-2}

\usepackage{graphicx}
\usepackage{dcolumn}
\usepackage{bm}
\usepackage{physics}
\usepackage{csquotes}
\usepackage{mathtools}
\usepackage{amsfonts}
\usepackage{amssymb}
\usepackage{amsmath}
\usepackage[caption=false,labelformat=empty]{subfig}
\usepackage{diagbox}
\usepackage{dsfont}
\usepackage{hyperref}
\usepackage{placeins}
\usepackage{xspace}
\usepackage{nameref}
\usepackage{orcidlink}
\usepackage[small]{complexity}

\usepackage{makecell}

\hypersetup{linktocpage,colorlinks,citecolor={blue},pdfdisplaydoctitle=true,pdfpagemode=UseOutlines,bookmarksnumbered=true}
\usepackage{verbatim}

\newcommand{\hodge}{{\star}}

\newcommand*{\id}{\ensuremath{\mathds{1}}}

\newcommand*{\gconfig}{\ensuremath{\mathcal{G}}}

\newcommand{\site}[1]{\ensuremath{\mathbf{#1}}}
\newcommand{\uvec}[1]{\ensuremath{\hat{\mathbf{#1}}}}
\newcommand{\N}{\ensuremath{\mathop{:}N\mathop{:}}}
\DeclarePairedDelimiter{\expectation}{\langle}{\rangle}

\newcommand{\ulm}{Institute for Complex Quantum Systems, Ulm University, 89069 Ulm, Germany}
\newcommand{\iqst}{Center for Integrated Quantum Science and Technology (IQST), Ulm-Stuttgart, Germany}
\newcommand{\racah}{Racah Institute of Physics, The Hebrew University of Jerusalem, Givat Ram, Jerusalem 91904, Israel}
\newcommand{\telaviv}{School of Physics and Astronomy, Tel Aviv University, Tel Aviv 6997801, Israel}

\begin{document}

\title{Sidestepping the Sign Problem with Tensor Networks for Lattice Gauge Theory}

\author{Ariel Kelman\,\orcidlink{0000-0002-7710-6538}}
\email{ariel.kelman@mail.huji.ac.il} 
\thanks{Corresponding author.}
\affiliation{\racah}
\affiliation{\telaviv}
\author{Gertian Roose\,\orcidlink{0009-0007-7608-0165}}
\affiliation{\telaviv}
\author{Patrick Emonts\,\orcidlink{0000-0002-7274-4071}}
\affiliation{\ulm}
\affiliation{\iqst}
\author{Erez Zohar\,\orcidlink{0000-0001-6993-6569}}
\affiliation{\telaviv}

\date{\today}

\begin{abstract}
Monte Carlo sampling of the Euclidean path integral is a powerful technique for computing static properties of lattice gauge theories, but it fails in the presence of a chemical potential for fermions and/or topological terms, where the Boltzmann weight becomes complex — the sign problem. Tensor-network-based variational approaches avoid the sign problem, give access to real-time dynamics, and can enforce gauge invariance exactly. However, contracting tensor network states in two or more spatial dimensions is exponentially costly unless one imposes further restrictions, such as Gaussianity, on the tensors. Gauged Gaussian Projected Entangled Pair States (GGPEPS) offer a hybrid method where the computation of expectation values reduces to sign-free Monte Carlo sampling over gauge-field configurations combined with Gaussian tensor network contractions over the matter fields, both of which scale polynomially with system size and dimension. Here we apply GGPEPS to a $\mathbb{Z}_2$ lattice gauge theory coupled to dynamical fermions at finite chemical potential in 2+1 dimensions, in a regime where conventional Monte Carlo suffers from a sign problem. Apart from the well-known confinement–deconfinement transition, we identify a phase at large chemical potential where string breaking happens in the presence of many background charges. These results provide the first demonstration that GGPEPS give a scalable route into sign-problem-afflicted regimes of lattice gauge theories.

\end{abstract}

\maketitle

\paragraph{Introduction} ---
\label{sec:introduction}
In recent history, the study of gauge theories has formed the backbone for a variety of fields. In high energy physics it forms the basis of the celebrated standard model of elementary particles and in condensed matter physics it has been used to effectively describe the low energy behavior of strongly interacting electrons~\cite{peskin_introduction_1995, Fradkin-FieldTheoriesCondensed-2013, Langacker-StandardModel-2017}. Furthermore, the presence of \enquote{stringy} degrees of freedom in gauge theories (i.e. Wilson lines) beautifully connects them to the flourishing field of higher symmetries~\cite{Gaiotto-GeneralizedGlobalSymmetries-2015, Gomes-IntroductionHigherformSymmetries-2023}. 

When gauge theories are weakly interacting, their properties can be studied perturbatively through the celebrated Feynman diagrams, and in this regime one finds great agreement with experiments~\cite{Gross-AsymptoticallyFreeGauge-1973}. However, when gauge theories are strongly interacting, for example in low-energy quantum chromodynamics, most modern analytical tools fall short and one has to rely on numerical tools, particularly in the framework of lattice gauge theories (LGTs)~\cite{Creutz-QuarksGluonsLattices-2023}.

Of these numerical tools, the Monte-Carlo sampling of observables with respect to the probability distribution $\propto \exp(-S_E)$ where $S_E$ is the Euclidean action has proven to be the most powerful one whenever it is applicable. In particular this approach has been used to obtain numerical estimates of the hadronic spectrum that agrees well with experiment~\cite{Aoki-FLAGReview2021-2022}. 
However, it is well known that this \enquote{probability distribution} becomes ill-defined due to taking on negative or complex values in the physically interesting regimes where the action includes a chemical potential or topological term --- a problem known as the \emph{sign problem}. 
Consequently numerical studies of gauge theories at high fermion densities, as is relevant for our understanding of systems as diverse as neutron stars and quark-gluon plasmas, currently remain out of reach of Monte Carlo. 

One approach to avoiding these challenges has been working with tensor networks (TNs)~\cite{cirac_renormalization_2009, orus_practical_2014, Cirac-MatrixProductStates-2021, Bridgeman-HandwavingInterpretiveDance-2017}. 
In 1D, tensor networks take the form of Matrix Product States (MPS), which have been studied extensively analytically and numerically. They have been used for ground state search together with algorithms such as DMRG~\cite{White-DensityMatrixFormulation-1992, White-DensitymatrixAlgorithmsQuantum-1993, schollwock_density-matrix_2011}, for studying time dynamics~\cite{daley_time-dependent_2004}, and for thermal states~\cite{verstraete_matrix_2004, zwolak_mixed-state_2004}.
They have further been applied to lattice gauge theories, in both particle and condensed matter physics, such as for determining mass spectra~\cite{Banuls-MassSpectrumSchwinger-2013}, ground states and excitations~\cite{Buyens-MatrixProductStates-2014}, and phase diagrams~\cite{tschirsich_phase_2019, Rico-TensorNetworksLattice-2014, gonzalez-cuadra_robust_2020, borla_quantum_2022}; 
general reviews can be found in~\cite{dalmonte_lattice_2016, banuls_simulating_2020, banuls_review_2020}. 

The generalization to higher dimensions can be approached in several ways, particularly into projected entangled pair states and its variants~\cite{cirac_renormalization_2009, Cirac-MatrixProductStates-2021, jordan_classical_2008, corboz_simulation_2010-1}.
However, in general the computational scaling of this approach is quite unfavorable~\cite{Verstraete-RenormalizationAlgorithmsQuantumMany-2004}.
Nevertheless they have been applied in a variety of settings, including lattice gauge theory where they have been particularly studied in connection with gauging mechanisms~\cite{haegeman_gauging_2015, zohar_building_2016} and Hamiltonian LGTs~\cite{tagliacozzo_entanglement_2011, Tagliacozzo-TensorNetworksLattice-2014, montangero_loop-free_2022}.
To avoid the high-dimensional complexity of tensor network contraction or capture the thermodynamic limit, one can modify the TN construction, such as in iPEPS and tree tensor networks~\cite{Felser-TwoDimensionalQuantumLinkLattice-2020,magnifico_lattice_2021, robaina_simulating_2021, cataldi_simulating_2024, Cataldi-RealTimeStringDynamics-2026}, or by  incorporating algorithms such as variational Monte Carlo~\cite{sorella_generalized_2001, sorella_wave_2005, bender_variational_2023, rouxinol_quantum_2026}; both of which provide a way of probing low-energy physics. 
The variational approach is also compatible with non-TN approaches, such as neural networks states~\cite{Banerjee-QuantumMonteCarlo-2024, Cohen-ComplexPathSimulations-2024, Spriggs-AccurateGroundStates-2025}.
Utilizing the gauging mechanisms just mentioned and combining them with fermionic PEPS~\cite{kraus_fermionic_2010}, \emph{gauged Gaussian projected entangled pair states} (GGPEPS) have been proposed as sign problem free hybrid of tensor networks and Monte-Carlo~\cite{zohar_fermionic_2015, zohar_projected_2016, Zohar-CombiningTensorNetworks-2018, Kelman-GaugedGaussianProjected-2024, Itou-AssessingExpressivePower-2026}. 
This ansatz is specialized for gauge theories, and preserves the gauge symmetry by construction. It also satisfies an entanglement area law, which is known or hypothesized to hold for ground states in gapped systems~\cite{hastings_area_2007, Eisert-ColloquiumAreaLaws-2010}, and is a general feature of tensor network states~\cite{orus_practical_2014, Cirac-MatrixProductStates-2021}.

In this work we apply GGPEPS to find the ground state of $\mathbb{Z}_2$ lattice gauge theories at finite fermion density in a regime where conventional Monte Carlo methods are known to suffer from a sign problem. The GGPEPS approach was previously used as a proof of concept in the pure gauge and matter settings without a chemical potential~\cite{emonts_finding_2023, kelman_projected_2026}. We start by reviewing the origin of the sign problem and summarizing the system and ansatz we study. We then analyze numerical results showing the ground state phase diagram.
In particular we find a weakly interacting phase, a strongly interacting phase with dynamical chiral symmetry breaking, and a phase at finite fermion density.

\paragraph{The Sign Problem} ---
\label{sec:sign_problem}
In standard Monte Carlo-based approaches to lattice gauge theory (or quantum many-bodied physics more generally), it is desired to, for example, compute the expectation value of an observable, 
\begin{equation} \begin{aligned}
    \expectation{\mathcal{O}}
        &= \frac{\int \mathcal{D}\phi \ \mathcal{O}(\phi) e^{-S_E(\phi)}}{\int \mathcal{D}\phi \ e^{-S_E(\phi)} } \\
\end{aligned} \end{equation}
where $\phi$ represents all the fields in the theory and $S_E(\phi)$ is the aforementioned Euclidean action, defined after a Wick rotation (which is necessary to get numerical integration techniques off the ground).
For the purposes of Monte Carlo, we define the probability distribution
\begin{equation} \begin{aligned}
\label{eq:prob}
    p(\phi)
        &= \frac{e^{-S_E(\phi)}}{\int \mathcal{D}\phi \ e^{-S_E(\phi)} } \\
\end{aligned} \end{equation}
though the value of the denominator is not needed, since the Metropolis sampling procedure only requires ratios of probabilities, in which the denominator cancels.
However, the expression $e^{-S_E(\phi)}$ is not guaranteed to be real and non-negative, despite that being a requirement in order for equation~\eqref{eq:prob} to be used as a probability distribution.

Suppose we have several fields, $\phi = \{ U, \bar{\psi}, \psi \}$, where 
the latter two are fermionic.
Then the normalization factor of the probability distribution, or equivalently the partition function, can be written as
\begin{equation} \begin{aligned}
    Z
        &= \int \mathcal{D}U \ \mathcal{D}\bar{\psi} \ \mathcal{D}\psi \ e^{-S_E(U, \bar{\psi}, \psi)} \\
        &= \int \mathcal{D}U \ e^{-S_G(U)} \det D(U),
\end{aligned} \end{equation}
where we have integrated out the fermionic fields in the second line, assuming $S_E(U, \bar{\psi}, \psi) = S_G(U) + S_F(U)$ where the second term is bilinear in the fermionic fields, and can be written as $S_F(U) = \bar{\psi} D(U) \psi$, for some operator $D(U)$ which depends on the bosonic fields $U$.

The condition mentioned above for avoiding the sign problem --- that $e^{-S_E(\phi)}$ be real and non-negative --- equivalently holds for $\det D(U)$.
If $\det D(U)$ is complex or negative for some $U$, then simulating the system will exhibit the sign problem; this happens, for example, when there is a non-zero chemical potential or when including a $\theta$-term on the field tensors
$(F_{\mu \nu} \ \hodge F^{\mu \nu} = \epsilon^{\mu \nu \alpha \beta} F_{\mu \nu} F_{\alpha \beta})$~\cite{Gattringer-QuantumChromodynamicsLattice-2010}. 
In the present work we consider a system with a finite chemical potential.
   
When the sign problem is present, one possible strategy is to define the probability into a magnitude and phase, $p(\phi) = \abs{p(\phi)} e^{i\Phi}$, and then absorb the phase into the observable $\mathcal{O}$. 
It can be shown that the relative error $\frac{ \delta \expectation{\mathcal{O}}}{\expectation{\mathcal{O}}}$ then scales as $ \frac{1/\sqrt{N}}{\abs{\expectation{ e^{i\Phi} }}}$, where $N$ is the number of samples.
In the thermodynamic limit, $\Phi$ scales with the volume, and so the phases will be approximately uniformly distributed on the unit circle, resulting in a diverging relative error. To compensate for this,  exponentially many samples are needed, which is the source of the computational difficulty of the sign problem.

The sign problem was first shown to be \NP-hard (for a particular system~\cite{Barahona-ComputationalComplexityIsing-1982}) in~\cite{Troyer-ComputationalComplexityFundamental-2005}. 
There it was argued that, since an analogous bosonic problem was efficiently simulate-able while the fermionic system was not, 
the source of the computational hardness is the fermionic statistics; the sign problem is often labeled the \textit{fermionic} sign problem for this reason.
In the present paper, we show how the GGPEPS ansatz allows us to efficiently and exactly satisfy the Gauss laws, an entanglement area law, and global symmetries, while working with variational Monte Carlo to sidestep the sign problem. This is not a fully general solution to the sign problem, but allows us to explore some theories in their sign-problem affected regimes.

\begin{figure}[th]
	\includegraphics[width=0.7\linewidth]{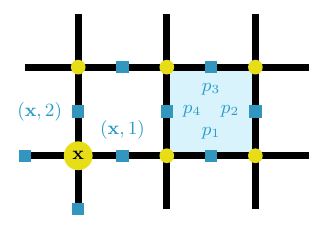}
	\caption{A 2D square lattice showing our labelling conventions. The fermionic matter is located on the lattice sites with gauge fields on the links.
    }
    \label{fig:system}
\end{figure}

\paragraph{System \& Ansatz} ---
\label{sec:ansatz}
We consider a system with a single flavor of fermions interacting with $\mathbb{Z}_2$ gauge fields on a square lattice with periodic boundary conditions, illustrated in figure~\ref{fig:system}.

For generic lattice gauge theories, the Hamiltonian with a lattice spacing $a=1$ is 
\begin{align}
H = \frac{\lambda}{2} H_\text{E} + \frac{1}{2\lambda} H_\text{B} + g_\text{I} H_\text{I} + m H_\text{M} + \mu \N,
\end{align}
where the couplings of the individual terms are all unitless. From left to right these terms represent the electric, magnetic, interaction, mass, and chemical terms respectively~\cite{Kogut-HamiltonianFormulationWilsons-1975}. The normal ordering of $N$ is taken with respect to the free massless fermion point which sits at half filling. 
For the $\mathbb{Z}_2$ lattice gauge theory specifically we have
\begin{equation} \begin{aligned}
    H_\text{E} &= \sum_\ell 2(1 - \sigma^z_\ell ), \\
    H_\text{B} &= \sum_{p} 2 (1 - \sigma^x_1 \sigma^x_2 \sigma^x_3 \sigma^x_4), \\
    H_\text{I}
        &= 
        \sum_{\site{x} }\Big( i \psi^\dagger(\site{x}) \sigma^x(\site{x},1) \psi(\site{x} + \uvec{e}_1) + h.c. \Big) \\
        & \ - \sum_{\site{x} }(-1)^{\site{x}} \Big( \psi^\dagger(\site{x}) \sigma^x(\site{x},2) \psi(\site{x} + \uvec{e}_2) + h.c. \Big), \\
    H_\text{M} &= \sum_{\site{x}} \Big((-1)^{\site{x}} \psi^\dagger(\site{x}) \psi(\site{x})\Big), \\
    \N &= \sum_{\site{x}} 
    \Big(
    \psi^\dagger(\site{x}) \psi(\site{x}) - \frac{1}{2}\Big),
\end{aligned} \end{equation} 
where $(-1)^\site{x} = (-1)^{x+y}$. 
The $(-1)^{\site{x}}$ factor arises from the staggering procedure, splitting matter and anti-matter onto even and odd sublattices~\cite{susskind_lattice_1977}. 
Crucially, this Hamiltonian is only defined on the the constrained Hilbert space where 
\begin{align}
    U_G(\site{x}) \ket{\psi} = e^{i \pi q_{\text{static}(\site{x})} } \ket{\psi} 
\end{align}
with 
\begin{equation} \begin{aligned}
    U_G(\site{x}) = e^{i \pi \psi^\dagger(\site{x}) \psi(\site{x}) } \sigma^z(\site{x}, 1)\sigma^z(\site{x}, 2)\sigma^z(\site{x}, -1)\sigma^z(\site{x}, -2),
\end{aligned} \end{equation}
valid for fixed static charges $q_\text{static}(\site{x}) \in \{ 0 , 1 \}$.

For a non-zero chemical potential, the system suffers from a sign problem --- that is, evaluating observables within given error using conventional Monte-Carlo is expected to have exponential cost as as a function of the desired error (see the previous section). 

To circumvent the sign problem and compute observables within a manageable timeframe we define a tensor network ansatz of the form 
\begin{equation} \begin{aligned}
    \label{eq:ansatz-integral-form}
    \ket{\Psi}
        &= \int \mathcal{D}\gconfig \ \ket{\psi(\gconfig)} \ket{\gconfig}, \\
\end{aligned} \end{equation}
where $\ket{\gconfig}$ are the possible gauge field configurations across the entire lattice, and $\ket{\psi(\gconfig)}$ is a corresponding state of the matter (satisfying the local gauge symmetry at each site). To construct the latter state we introduce \enquote{virtual} modes associated with each lattice site, and with each direction (positive and negative) of the lattice.
The state is then constructed as
\begin{equation} \begin{aligned}
    \label{eq:ansatz-for-gauge-config}
    \ket{\psi(\gconfig)}
        &= \bra{\Omega_v} \prod_\ell \omega(l) \prod_\ell \mathcal{U}^\gconfig(l) \prod_{\site{x}} A(\site{x}) \ket{\Omega_v} \ket{\Omega_p} \\
\end{aligned} \end{equation}
where 
\begin{equation} \begin{aligned}
    \label{eq:ansatz-components}
    A(\site{x}) &= \exp \Big( \mathcal{T}_{ij}(\site{x}) 
        \alpha^\dagger_i(\site{x}) \alpha^\dagger_j(\site{x}) \Big), \\
    \mathcal{U}^\gconfig(\site{x}, k) &=  \id_q \otimes \mathds{P}_{\textnormal{even}}(\site{x}, k) + \sigma^x(\site{x}, k) \otimes \mathds{P}_{\textnormal{odd}}(\site{x}, k) \\
    \omega(\site{x}, k) &= \exp \Big( W^{(k)}_{ij}
    \alpha^{\dagger}_i(\site{x})
    \alpha^{\dagger}_j(\site{x} + \site{e}_k)
    \Big).
\end{aligned} \end{equation}
We denote all fermionic creation operators by $\alpha_i^\dagger$, which include both the physical modes and the additional virtual modes used to define the tensor network, whose vacuum states are $\ket{\Omega_p}, \ket{\Omega_v}$ respectively. 
In the $A(\site{x})$ operator, $i, j$ range over all physical modes associated with site $\site{x}$; in $\omega(\site{x}, k)$ they range over the virtual modes associated with the link $\ell = (\site{x}, k)$.
The complex matrices $\mathcal{T}$ and $W^{(k)}$ can be freely chosen, so long as the choice is compatible with the desired symmetries. 
We allow the parameters in $\mathcal{T}$ to vary to define different states.
The projectors $\mathds{P}_{\textnormal{even/odd}}$ project onto the subspaces of even/odd filling of the virtual modes on the given link.

Conceptually, the state is constructed in three stages, corresponding to the three products of equation~\eqref{eq:ansatz-for-gauge-config}: an initial stage which couples the fermionic matter on each site with virtual modes, one per neighboring link; a stage which couples the gauge fields on each link to the aforementioned virtual modes and protects the gauge symmetry; and a final stage which entangles neighboring virtual modes and then traces them out.
We have argued elsewhere that this form satisfies many desirable qualities in an ansatz~\cite{Kelman-GaugedGaussianProjected-2024, kelman_projected_2026}; here we simply use it to demonstrate results in a sign problem affected regime.

We wish to be very clear about our claim: we are not claiming to have solved the sign problem, that is, to have given an efficient (polynomial in the number of degrees of freedom) algorithm for arbitrary calculations suffering from the sign problem; that would imply a significant change to our understanding of complexity theory. 
Rather, by constructing a specially built ansatz to be used with variational Monte Carlo, we have sacrificed general expressibility for avoiding the exponential cost of the sign problem in particular situations. 
Simultaneously, our ansatz satisfies other important desiderata --- it satisfies an area (boundary) entanglement law, and has a local gauge symmetry --- making it useful for studying systems, such as LGTs, with which other methods struggle.
In the following section, we show results for a lattice up to size $6 \times 6$, larger than what can easily be handled through other methods.

The GGPEPS ansatz trades some expressibility for computational tractability. The limitation on expressibility arises from the Gaussian nature of the individual terms making up the state $\ket{\psi(\gconfig)}$ (though the state $\ket{\Psi}$ is non-Gaussian when considered in totality) --- it is designed for capturing low energy states of lattice gauge theory. 
In particular, when the gauge interactions are very weak or very strong the ground state is expected to work very well, at the transition there is no a priori guarantee but we observe consistent results. 
We note that the ansatz can be expanded by gauging superpositions of Gaussian PEPS which expands the expressibility at the expense of computational cost~\cite{Roose-GaugingSuperpositionFermionic-2025}.

\begin{figure}[h]
	\includegraphics[width=0.98\linewidth]{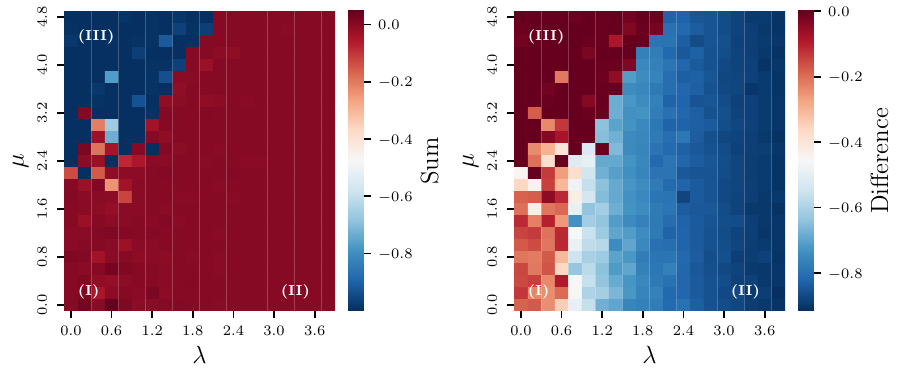} 
	\caption{The expectation values of the total number of particles in the ground state (left), the order parameter for the breaking of translation symmetry $N_\text{even} - N_\text{odd}$ (right). The mass and hopping strength are fixed to $m=0.2$ and $g_I = 1.0$ respectively, the system size was $6 \times 6$. 
    The Roman numerals label the phases discussed in the main text.} 
    \label{fig:results_phase}
\end{figure}

\paragraph{Results \& Discussion} ---
\label{sec:results}
We obtained ground state estimates of the $\mathbb{Z}_2$ lattice gauge theory Hamiltonian for various values of the gauge coupling $\lambda$ and chemical potential $\mu$ while keeping $g_I = 1.0$ and $m = 0.2$ fixed. 
We show results for a system size of $6 \times 6$; this system has Hilbert space dimension $2^{73}$ after accounting for the gauge symmetry constraints. 
We found an interesting phase diagram that contains (I) a weakly interacting phase, (II) a strongly interacting confined phase with dynamically generated massive particle particles, and (III) a phase at nontrivial particle density. 
In figure~\ref{fig:results_phase} these phases respectively appear in the top-left, right and bottom-left respectively. In what follows we will explain these claims in more detail through the introduction of suitable order parameters and other phase indicators.  

We first focus on the number density. Since $[H, \N] = 0$, eigenstates of the Hamiltonian can be distinguished by the number of particles (or holes) that they contain compared to the half filled state. In the language of symmetries, $\langle\N\rangle$ can be thought of as the irrep of the groundstate with respect to the fermionic vector symmetry
\begin{align}
    U_\text{vector}(\theta) = \exp \Big( i \theta \sum_\site{x} \psi^\dagger(\site{x}) \psi(\site{x}) \Big).
\end{align}
In the left panel of figure~\ref{fig:results_phase} we clearly see that the ground state for large $\mu$ (i.e. top-left) has negative $\expval{\N}$ so that this state contains a finite density of holes compared to the remainder of the phase diagram. Furthermore, we observe that the phase boundaries of the finite density phase are sharp i.e. the transition is first order. This makes sense because the spectrum of \N is quantized and all groundstates are exact eigenstates of $\N$ (since $[H, \N]=0$).

In 1+1 dimensional lattice field theory it is known that lattice translations of staggered fermions implement the $\mathbb{Z}_2$ subgroup of the continuum chiral symmetry, therefore the breaking of lattice translations is naturally associated with mass generation~\cite{susskind_lattice_1977, Kogut-HamiltonianFormulationWilsons-1975}. 
In 2+1 dimensions the situation is more complicated because chirality is not defined for 2+1 dimensional fermions in the fundamental representation of the Lorentz group. 
However, the continuum limit of free staggered 2+1 dimensional lattice fermions describes two Dirac cones and therefore the natural field theory description of our lattice model lives in the $4 = 2 \bigoplus 2$ dimensional representation of the Lorentz group (i.e. \emph{one fundamental 2 dimensional Weyl fermion per Dirac cone}). 
In this representation there are 2 independent well-defined chiral symmetries that correspond to the two lattice translations
\begin{align}
    \psi^\dagger(\site{x}) \rightarrow (-1)^\site{x} \psi^\dagger(\site{x} + \site{e}_k).
\end{align}
Therefore, as in the one dimensional case it is natural to interpret breaking of translation symmetry as the breaking of chiral symmetry which signals (possibly dynamical) mass generation. 
In the right panel of figure~\ref{fig:results_phase} this chiral symmetry breaking is analyzed through the expectation value of the associated order parameter $\Delta N = N_\text{even} - N_\text{odd}$, i.e. the difference in the expectation value of the number operator on even and odd sites; we identify three distinct phases. 
In the weakly interacting phase (bottom-left) we find a small nonzero amount of chiral symmetry breaking as a consequence of the chosen mass $m=0.2$. 
For large gauge interactions $\lambda$ where the gauge fields confine the matter we find a large increase in magnitude of $\Delta N$. 
The associated mass appears dynamical through a non-perturbative process first described in the Gross Neveu model~\cite{Gross-DynamicalSymmetryBreaking-1974}; see~\cite{Roose-LatticeRegularisationEntanglement-2021, Pannullo-LatticeInvestigationPhase-2020} for a detailed TN analysis of this model. 
Finally, for large chemical potential i.e. in the phase with nontrivial particle number, we find that the translation symmetry is present. 

\begin{figure}[h]
	\includegraphics[width=0.5\linewidth]{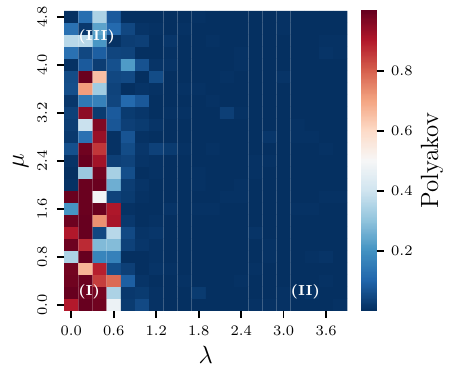} \\
	\caption{The expectation values of (the absolute value of) a Polyakov loop around the torus, for the system with $m=0.2$ and $g_I = 1.0$; the system size was $6 \times 6$.} 
    \label{fig:results_phase_polyakov}
\end{figure}

Another interesting observable is the Polyakov loop
\begin{align}
    \langle W_\text{Pol.} \rangle = \langle \Pi_{l \in C} X_\ell \rangle 
\end{align}
where $C$ is a contour chosen to wrap around the torus~\cite{Polyakov-CompactGaugeFields-1975}. Physically nonzero $\langle W_\text{Pol.} \rangle$ signals that the ground state contains Wilson lines that are as large as the system size. 
Therefore it is expected that $\langle W_\text{Pol.} \rangle$ is zero in the strongly interacting confining phase and nonzero in the weakly interacting phase. In figure~\ref{fig:results_phase_polyakov} this is indeed what we observe, though the ansatz shows more noise for this sort of long-range observable which does not appear in the Hamiltonian. Additionally, we see that the value of $\langle W_\text{top.} \rangle$ is zero in the finite density phase. This suggests that the finite charge density leads to string breaking. Finally, note that this quantity is not an order parameter for generic values of the couplings because it is not the charged object with respect to some symmetry. Therefore, generic phase transitions between the weakly interacting and confined regime are not expected to be second order. It is however interesting to note the special case $g_I = 0$ (i.e. no interactions with fermions) where the expectation value of $\langle W_\text{Pol.} \rangle$ becomes an order parameter for the breaking of the one-form symmetry 
$\prod_{l \in \tilde C} Z_\ell$
where $\tilde C$ is a closed loop on the dual lattice~\cite{Gaiotto-GeneralizedGlobalSymmetries-2015, Gomes-IntroductionHigherformSymmetries-2023}.

\begin{figure}[h]
	\includegraphics[width=0.9\linewidth]{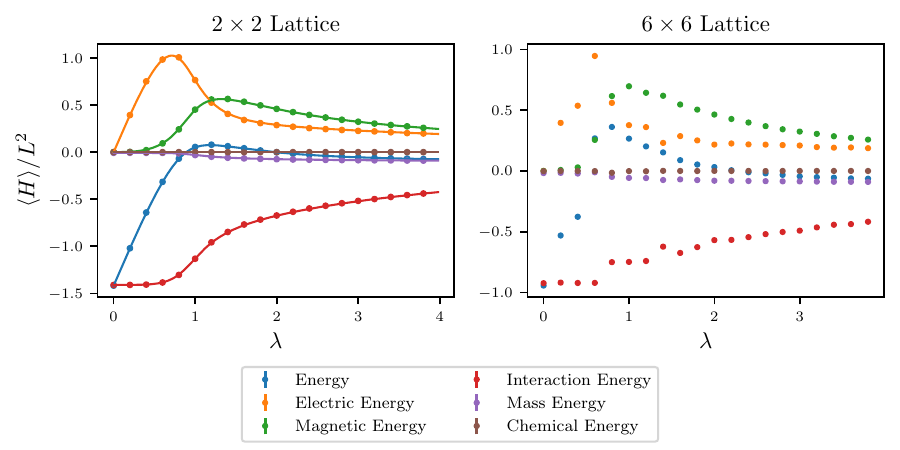}
	\caption{A scan of the normalized ground state energy terms, at $g_I = 1.0, m = 0.2, \mu = 1.0$. The leftmost plot shows results computed by explicitly summing over the $2^8$ gauge configurations of a $2 \times 2$ lattice; the $6\times 6$ plot was computed using Monte Carlo (the Monte Carlo errors shown on the plot are quite small; we have not quantified error due to the minimization procedure). The $2 \times 2$ results are compared with results computed by exactly diagonalizing the Hamiltonian (\emph{solid lines}).}
    \label{fig:results}
\end{figure}

To end this section we comment on the quality of these numerical estimates. To do so, we first focus on a small $2 \times 2$ lattice where it is possible to exactly diagonalize the Hamiltonian. Because the number of possible gauge configurations on a $2\times 2$ lattice is relatively small ($2^8 = 256$) we can evaluate the sum in equation~\eqref{eq:ansatz-integral-form} exactly. In the leftmost panel of figure~\ref{fig:results} we explicitly compared this exact contraction of equation $\eqref{eq:ansatz-integral-form}$ (dots) with exact diagonalization (lines) and found great agreement even across the phase transition. 
In the remaining panel we increase the system size and find good similar  behavior to the first panel, indicating that the data remains trustworthy for larger system sizes (note that as shown in~\cite{kelman_projected_2026}, the interaction energy is not proportional to lattice size for small lattices). 

\paragraph{Conclusion} ---
\label{sec:conclusions}
We have used gauged Gaussian projected entangled pair states (GGPEPS) to numerically study a $\mathbb{Z}_2$ lattice gauge theory with dynamical fermions at large chemical potential. Notably, this region of the phase diagram is inaccessible to Monte Carlo integration due to the sign problem. This result firmly establishes GGPEPS as a numerical tool well suited for obtaining physically relevant results. Owing to its favorable scaling to 3+1 dimensions and to other gauge groups, we hope that this opens the door to eventually studying quantum chromodynamics at finite density.

Apart from this numerical progress, we also carefully analyzed the obtained phase diagram in the language of generalized (higher-form) symmetry breaking. In particular, we found an interesting phase at large chemical potential where the high density of charged particles leads to string breaking of the Wilson lines.

In future work, we will apply this method to finite non-Abelian groups such as $D_N$, and take a first step towards Lie groups by studying $\mathbb{Z}_N$ in the limit of large $N$.

\paragraph{Data \& Code Availability} ---
\label{sec:data_availability}
All data is available at the Zenodo repository~\cite{zenodo_23889660}.
The \texttt{ggpeps} code is available at~\cite{Kelman-GGPEPSGithubSource-2026}, and described in~\cite{Kelman-GgpepsPackageGauged-2026}.

\paragraph{Acknowledgments} ---
\label{sec:acknowledgments}
We would like to thank U. Borla, J. Bender, I. Gomelski, and J. Elyovich for helpful discussions.
The results were computed using HPC resources at Tel Aviv University.
AI was used in the coding workflow, with manual verification, and extensive unit testing. AI was also used to help refine the paper’s wording.
This work was supported by the European Union Research Council (ERC, OverSign, 101122583). Views and opinions expressed are those of the authors only and do not necessarily reflect those of the European Union or the European Research Council.
This work was also supported through the NWO-Quantum Technology program (Grant No.~NGF.1623.23.006, DOI:~\url{https://doi.org/10.61686/RBLDP65717}) and funding by the Carl-Zeiss-Stiftung (CZS Center QPhoton).

\bibliography{references_bibtex}

@article{Aoki-FLAGReview2021-2022,
  title = {{{FLAG Review}} 2021},
  author = {Aoki, Y. and Blum, T. and Colangelo, G. and Collins, S. and Morte, M. Della and Dimopoulos, P. and D{\"u}rr, S. and Feng, X. and Fukaya, H. and Golterman, M. and Gottlieb, Steven and Gupta, R. and Hashimoto, S. and Heller, U. M. and Herdoiza, G. and Hernandez, P. and Horsley, R. and J{\"u}ttner, A. and Kaneko, T. and Lunghi, E. and Meinel, S. and Monahan, C. and Nicholson, A. and Onogi, T. and Pena, C. and Petreczky, P. and Portelli, A. and Ramos, A. and Sharpe, S. R. and Simone, J. N. and Simula, S. and Sint, S. and Sommer, R. and Tantalo, N. and {Van de Water}, R. and Wenger, U. and Wittig, H.},
  year = 2022,
  month = oct,
  journal = {The European Physical Journal C},
  volume = {82},
  number = {10},
  pages = {1--296},
  publisher = {Springer Berlin Heidelberg},
  issn = {1434-6052},
  doi = {10.1140/epjc/s10052-022-10536-1},
  urldate = {2023-03-22},
  copyright = {2022 Crown},
  langid = {english}
}

@article{Banerjee-QuantumMonteCarlo-2024,
  title = {Quantum {{Monte Carlo}} for Gauge Fields and Matter without the Fermion Determinant},
  author = {Banerjee, Debasish and Huffman, Emilie},
  year = 2024,
  month = feb,
  journal = {Physical Review D},
  volume = {109},
  number = {3},
  pages = {L031506},
  publisher = {American Physical Society},
  doi = {10.1103/PhysRevD.109.L031506},
  urldate = {2026-11-19}
}

@article{banuls_review_2020,
  ids = {banuls_review_2019,carmen_banuls_review_2020},
  title = {Review on Novel Methods for Lattice Gauge Theories},
  author = {Ba{\~n}uls, Mari Carmen and Cichy, Krzysztof},
  year = 2020,
  month = feb,
  journal = {Reports on Progress in Physics},
  volume = {83},
  number = {2},
  pages = {024401},
  issn = {0034-4885, 1361-6633},
  doi = {10.1088/1361-6633/ab6311},
  urldate = {2020-07-30}
}

@article{banuls_simulating_2020,
  title = {Simulating Lattice Gauge Theories within Quantum Technologies},
  author = {Ba{\~n}uls, Mari Carmen and Blatt, Rainer and Catani, Jacopo and Celi, Alessio and Cirac, Juan Ignacio and Dalmonte, Marcello and Fallani, Leonardo and Jansen, Karl and Lewenstein, Maciej and Montangero, Simone and Muschik, Christine A. and Reznik, Benni and Rico, Enrique and Tagliacozzo, Luca and Van Acoleyen, Karel and Verstraete, Frank and Wiese, Uwe-Jens and Wingate, Matthew and Zakrzewski, Jakub and Zoller, Peter},
  year = 2020,
  month = aug,
  journal = {The European Physical Journal D},
  volume = {74},
  number = {8},
  pages = {165},
  issn = {1434-6060, 1434-6079},
  doi = {10.1140/epjd/e2020-100571-8},
  urldate = {2022-03-10},
  langid = {english}
}

@article{Banuls-MassSpectrumSchwinger-2013,
  title = {The Mass Spectrum of the {{Schwinger}} Model with Matrix Product States},
  author = {Ba{\~n}uls, M.C. and Cichy, K. and Cirac, J.I. and Jansen, K.},
  year = 2013,
  month = nov,
  journal = {Journal of High Energy Physics},
  volume = {2013},
  number = {11},
  pages = {158},
  issn = {1029-8479},
  doi = {10.1007/JHEP11(2013)158},
  urldate = {2026-11-22},
  langid = {english}
}

@article{Barahona-ComputationalComplexityIsing-1982,
  title = {On the Computational Complexity of {{Ising}} Spin Glass Models},
  author = {Barahona, F.},
  year = 1982,
  month = oct,
  journal = {Journal of Physics A: Mathematical and General},
  volume = {15},
  number = {10},
  pages = {3241},
  issn = {0305-4470},
  doi = {10.1088/0305-4470/15/10/028},
  urldate = {2025-12-09},
  langid = {english}
}

@article{bender_variational_2023,
  title = {Variational {{Monte Carlo}} Algorithm for Lattice Gauge Theories with Continuous Gauge Groups: {{A}} Study of ( 2 + 1 ) -Dimensional Compact {{QED}} with Dynamical Fermions at Finite Density},
  shorttitle = {Variational {{Monte Carlo}} Algorithm for Lattice Gauge Theories with Continuous Gauge Groups},
  author = {Bender, Julian and Emonts, Patrick and Cirac, J. Ignacio},
  year = 2023,
  month = nov,
  journal = {Physical Review Research},
  volume = {5},
  number = {4},
  pages = {043128},
  issn = {2643-1564},
  doi = {10.1103/PhysRevResearch.5.043128},
  urldate = {2023-11-14},
  langid = {english}
}

@article{borla_quantum_2022,
  title = {Quantum Phases of Two-Dimensional {{Z}}{$_2$} Gauge Theory Coupled to Single-Component Fermion Matter},
  author = {Borla, Umberto and Jeevanesan, Bhilahari and Pollmann, Frank and Moroz, Sergej},
  year = 2022,
  month = feb,
  journal = {Physical Review B},
  volume = {105},
  number = {7},
  pages = {075132},
  issn = {2469-9950, 2469-9969},
  doi = {10.1103/PhysRevB.105.075132},
  urldate = {2022-10-18}
}

@article{Bridgeman-HandwavingInterpretiveDance-2017,
  title = {Hand-Waving and {{Interpretive Dance}}: {{An Introductory Course}} on {{Tensor Networks}}},
  shorttitle = {Hand-Waving and {{Interpretive Dance}}},
  author = {Bridgeman, Jacob C. and Chubb, Christopher T.},
  year = 2017,
  month = jun,
  journal = {Journal of Physics A: Mathematical and Theoretical},
  volume = {50},
  number = {22},
  eprint = {1603.03039},
  primaryclass = {cond-mat, physics:hep-th, physics:quant-ph},
  pages = {223001},
  issn = {1751-8113, 1751-8121},
  doi = {10.1088/1751-8121/aa6dc3},
  urldate = {2022-06-23},
  archiveprefix = {arXiv}
}

@article{Buyens-MatrixProductStates-2014,
  title = {Matrix {{Product States}} for {{Gauge Field Theories}}},
  author = {Buyens, Boye and Haegeman, Jutho and Van Acoleyen, Karel and Verschelde, Henri and Verstraete, Frank},
  year = 2014,
  month = aug,
  journal = {Physical Review Letters},
  volume = {113},
  number = {9},
  pages = {091601},
  publisher = {American Physical Society},
  doi = {10.1103/PhysRevLett.113.091601},
  urldate = {2026-11-22}
}

@article{cataldi_simulating_2024,
  title = {Simulating (2+1){{D SU}}(2) {{Yang-Mills}} Lattice Gauge Theory at Finite Density with Tensor Networks},
  author = {Cataldi, Giovanni and Magnifico, Giuseppe and Silvi, Pietro and Montangero, Simone},
  year = 2024,
  month = jul,
  journal = {Physical Review Research},
  volume = {6},
  number = {3},
  pages = {033057},
  publisher = {American Physical Society},
  doi = {10.1103/PhysRevResearch.6.033057}
}

@article{Cataldi-RealTimeStringDynamics-2026,
  title = {Real-{{Time String Dynamics}} in a \$2+1\textbackslash mathrm\textbraceleft{{D}}\textbraceright\$ {{Non-Abelian Lattice Gauge Theory}}: {{String Breaking}}, {{Glueball Formation}}, {{Baryon Blockade}}, and {{Tension Reduction}}},
  shorttitle = {Real-{{Time String Dynamics}} in a \$2+1\textbackslash mathrm\textbraceleft{{D}}\textbraceright\$ {{Non-Abelian Lattice Gauge Theory}}},
  author = {Cataldi, Giovanni and Orlando, Simone and Halimeh, Jad C.},
  year = 2026,
  month = oct,
  journal = {Physical Review Letters},
  volume = {137},
  number = {16},
  pages = {161902},
  publisher = {American Physical Society},
  doi = {10.1103/52gk-hh7y},
  urldate = {2026-11-19}
}

@article{cirac_renormalization_2009,
  ids = {cirac_renormalization_2009-1},
  title = {Renormalization and Tensor Product States in Spin Chains and Lattices},
  author = {Cirac, J. I. and Verstraete, F.},
  year = 2009,
  journal = {Journal of Physics A: Mathematical and Theoretical},
  volume = {42},
  number = {50},
  pages = {504004},
  doi = {10.1088/1751-8113/42/50/504004}
}

@article{Cirac-MatrixProductStates-2021,
  title = {Matrix Product States and Projected Entangled Pair States: {{Concepts}}, Symmetries, Theorems},
  shorttitle = {Matrix Product States and Projected Entangled Pair States},
  author = {Cirac, J. Ignacio and {P{\'e}rez-Garc{\'i}a}, David and Schuch, Norbert and Verstraete, Frank},
  year = 2021,
  month = dec,
  journal = {Reviews of Modern Physics},
  volume = {93},
  number = {4},
  pages = {045003},
  publisher = {American Physical Society},
  doi = {10.1103/RevModPhys.93.045003},
  urldate = {2023-03-23}
}

@article{Cohen-ComplexPathSimulations-2024,
  title = {Complex Path Simulations of Geometrically Frustrated Ladders},
  author = {Cohen, Elyasaf Y. and Alexandru, Andrei and Gazit, Snir},
  year = 2024,
  month = mar,
  journal = {Physical Review B},
  volume = {109},
  number = {11},
  pages = {115122},
  publisher = {American Physical Society},
  doi = {10.1103/PhysRevB.109.115122},
  urldate = {2026-11-19}
}

@article{corboz_simulation_2010-1,
  ids = {corboz_simulation_2010-2},
  title = {Simulation of Strongly Correlated Fermions in Two Spatial Dimensions with Fermionic Projected Entangled-Pair States},
  author = {Corboz, Philippe and Or{\'u}s, Rom{\'a}n and Bauer, Bela and Vidal, Guifr{\'e}},
  year = 2010,
  month = apr,
  journal = {Physical Review B},
  volume = {81},
  number = {16},
  pages = {165104},
  publisher = {American Physical Society},
  doi = {10.1103/PhysRevB.81.165104}
}

@book{Creutz-QuarksGluonsLattices-2023,
  title = {Quarks, {{Gluons}} and {{Lattices}}},
  author = {Creutz, Michael},
  year = 2023,
  series = {Cambridge {{Monographs}} on {{Mathematical Physics}}},
  publisher = {Cambridge University Press},
  address = {Cambridge},
  doi = {10.1017/9781009290395},
  urldate = {2026-06-18},
  isbn = {978-1-009-29038-8}
}

@article{daley_time-dependent_2004,
  ids = {daley_time-dependent_2004-1},
  title = {Time-Dependent Density-Matrix Renormalization-Group Using Adaptive Effective {{Hilbert}} Spaces},
  author = {Daley, A J and Kollath, C and Schollw{\"o}ck, U and Vidal, G},
  year = 2004,
  month = apr,
  journal = {Journal of Statistical Mechanics: Theory and Experiment},
  volume = {2004},
  number = {04},
  pages = {P04005},
  issn = {1742-5468},
  doi = {10.1088/1742-5468/2004/04/P04005},
  urldate = {2020-06-15}
}

@article{dalmonte_lattice_2016,
  title = {Lattice Gauge Theory Simulations in the Quantum Information Era},
  author = {Dalmonte, M. and Montangero, S.},
  year = 2016,
  month = jul,
  journal = {Contemporary Physics},
  volume = {57},
  number = {3},
  pages = {388--412},
  issn = {0010-7514, 1366-5812},
  doi = {10.1080/00107514.2016.1151199},
  urldate = {2019-06-19},
  langid = {english}
}

@article{Eisert-ColloquiumAreaLaws-2010,
  title = {Colloquium: {{Area}} Laws for the Entanglement Entropy},
  shorttitle = {Colloquium},
  author = {Eisert, J. and Cramer, M. and Plenio, M. B.},
  year = 2010,
  month = feb,
  journal = {Reviews of Modern Physics},
  volume = {82},
  number = {1},
  pages = {277--306},
  publisher = {American Physical Society},
  doi = {10.1103/RevModPhys.82.277},
  urldate = {2023-03-28}
}

@article{emonts_finding_2023,
  title = {Finding the Ground State of a Lattice Gauge Theory with Fermionic Tensor Networks: A $2+1d$ $\mathbb{Z}_2$ Demonstration},
  shorttitle = {Finding the Ground State of a Lattice Gauge Theory with Fermionic Tensor Networks},
  author = {Emonts, Patrick and Kelman, Ariel and Borla, Umberto and Moroz, Sergej and Gazit, Snir and Zohar, Erez},
  year = 2023,
  month = jan,
  journal = {Physical Review D},
  volume = {107},
  number = {1},
  eprint = {2211.00023},
  primaryclass = {cond-mat, physics:hep-lat, physics:quant-ph},
  pages = {014505},
  issn = {2470-0010, 2470-0029},
  doi = {10.1103/PhysRevD.107.014505},
  urldate = {2023-02-28},
  archiveprefix = {arXiv}
}

@article{Felser-TwoDimensionalQuantumLinkLattice-2020,
  title = {Two-{{Dimensional Quantum-Link Lattice Quantum Electrodynamics}} at {{Finite Density}}},
  author = {Felser, Timo and Silvi, Pietro and Collura, Mario and Montangero, Simone},
  year = 2020,
  month = nov,
  journal = {Physical Review X},
  volume = {10},
  number = {4},
  pages = {041040},
  publisher = {American Physical Society},
  doi = {10.1103/PhysRevX.10.041040},
  urldate = {2026-06-10}
}

@book{Fradkin-FieldTheoriesCondensed-2013,
  title = {Field {{Theories}} of {{Condensed Matter Physics}}},
  author = {Fradkin, Eduardo},
  year = 2013,
  month = apr,
  edition = {2},
  publisher = {Cambridge University Press}
}

@article{Gaiotto-GeneralizedGlobalSymmetries-2015,
  title = {Generalized Global Symmetries},
  author = {Gaiotto, Davide and Kapustin, Anton and Seiberg, Nathan and Willett, Brian},
  year = 2015,
  month = feb,
  journal = {Journal of High Energy Physics},
  volume = {2015},
  number = {2},
  pages = {172},
  issn = {1029-8479},
  doi = {10.1007/JHEP02(2015)172},
  urldate = {2026-11-19},
  langid = {english}
}

@book{Gattringer-QuantumChromodynamicsLattice-2010,
  title = {Quantum {{Chromodynamics}} on the {{Lattice}}: {{An Introductory Presentation}}},
  shorttitle = {Quantum {{Chromodynamics}} on the {{Lattice}}},
  author = {Gattringer, Christof and Lang, Christian B.},
  year = 2010,
  series = {Lecture {{Notes}} in {{Physics}}},
  volume = {788},
  publisher = {Springer},
  address = {Berlin, Heidelberg},
  doi = {10.1007/978-3-642-01850-3},
  urldate = {2025-11-12},
  copyright = {https://www.springernature.com/gp/researchers/text-and-data-mining},
  isbn = {978-3-642-01849-7 978-3-642-01850-3},
  langid = {english}
}

@article{Gomes-IntroductionHigherformSymmetries-2023,
  title = {An Introduction to Higher-Form Symmetries},
  author = {Gomes, Pedro R. S.},
  year = 2023,
  month = sep,
  journal = {SciPost Physics Lecture Notes},
  pages = {074},
  issn = {2590-1990},
  doi = {10.21468/SciPostPhysLectNotes.74},
  urldate = {2026-11-19},
  langid = {english}
}

@article{gonzalez-cuadra_robust_2020,
  title = {Robust {{Topological Order}} in {{Fermionic Z}}{$_2$} {{Gauge Theories}}: {{From Aharonov-Bohm Instability}} to {{Soliton-Induced Deconfinement}}},
  shorttitle = {Robust {{Topological Order}} in {{Fermionic}} \$\textbraceleft\textbackslash mathbb\textbraceleft{{Z}}\textbraceright\textbraceright\_\textbraceleft 2\textbraceright\$ {{Gauge Theories}}},
  author = {{Gonz{\'a}lez-Cuadra}, Daniel and Tagliacozzo, Luca and Lewenstein, Maciej and Bermudez, Alejandro},
  year = 2020,
  month = oct,
  journal = {Physical Review X},
  volume = {10},
  number = {4},
  pages = {041007},
  publisher = {American Physical Society},
  doi = {10.1103/PhysRevX.10.041007},
  urldate = {2022-10-24}
}

@article{Gross-AsymptoticallyFreeGauge-1973,
  title = {Asymptotically {{Free Gauge Theories}}. {{I}}},
  author = {Gross, David J. and Wilczek, Frank},
  year = 1973,
  month = nov,
  journal = {Physical Review D},
  volume = {8},
  number = {10},
  pages = {3633--3652},
  publisher = {American Physical Society},
  doi = {10.1103/PhysRevD.8.3633},
  urldate = {2023-03-22}
}

@article{Gross-DynamicalSymmetryBreaking-1974,
  title = {Dynamical Symmetry Breaking in Asymptotically Free Field Theories},
  author = {Gross, David J. and Neveu, Andr{\'e}},
  year = 1974,
  month = nov,
  journal = {Physical Review D},
  volume = {10},
  number = {10},
  pages = {3235--3253},
  publisher = {American Physical Society},
  doi = {10.1103/PhysRevD.10.3235},
  urldate = {2026-11-12}
}

@article{haegeman_gauging_2015,
  title = {Gauging {{Quantum States}}: {{From Global}} to {{Local Symmetries}} in {{Many-Body Systems}}},
  shorttitle = {Gauging {{Quantum States}}},
  author = {Haegeman, Jutho and Van Acoleyen, Karel and Schuch, Norbert and Cirac, J. Ignacio and Verstraete, Frank},
  year = 2015,
  month = feb,
  journal = {Physical Review X},
  volume = {5},
  number = {1},
  pages = {011024},
  publisher = {American Physical Society (APS)},
  issn = {2160-3308},
  doi = {10.1103/PhysRevX.5.011024},
  urldate = {2022-11-14},
  langid = {english}
}

@article{hastings_area_2007,
  title = {An Area Law for One-Dimensional Quantum Systems},
  author = {Hastings, M B},
  year = 2007,
  journal = {Journal of Statistical Mechanics: Theory and Experiment},
  volume = {2007},
  number = {08},
  eprint = {0705.2024},
  pages = {P08024},
  archiveprefix = {arXiv}
}

@misc{Itou-AssessingExpressivePower-2026,
  title = {Assessing the Expressive Power of a Gauge-Invariant {{PEPS}} Ansatz across a Topological Phase Transition},
  author = {Itou, Etsuko and Okubo, Tsuyoshi},
  year = 2026,
  month = nov,
  number = {arXiv:2611.13852},
  eprint = {2611.13852},
  primaryclass = {hep-lat},
  publisher = {arXiv},
  doi = {10.48550/arXiv.2611.13852},
  urldate = {2026-11-22},
  archiveprefix = {arXiv}
}

@article{jordan_classical_2008,
  title = {Classical {{Simulation}} of {{Infinite-Size Quantum Lattice Systems}} in {{Two Spatial Dimensions}}},
  author = {Jordan, J. and Or{\'u}s, R. and Vidal, G. and Verstraete, F. and Cirac, J. I.},
  year = 2008,
  month = dec,
  journal = {Physical Review Letters},
  volume = {101},
  number = {25},
  pages = {250602},
  publisher = {American Physical Society},
  doi = {10.1103/PhysRevLett.101.250602},
  urldate = {2022-04-06}
}

@article{kelman_projected_2026,
  title = {Projected Entangled Pair States for Lattice Gauge Theories with Dynamical Fermions},
  author = {Kelman, Ariel and Borla, Umberto and Emonts, Patrick and Zohar, Erez},
  year = 2026,
  month = jan,
  journal = {Communications Physics},
  publisher = {Nature Publishing Group},
  issn = {2399-3650},
  doi = {10.1038/s42005-025-02482-7},
  urldate = {2026-01-15},
  copyright = {2026 The Author(s)},
  langid = {english}
}

@article{Kelman-GaugedGaussianProjected-2024,
  title = {Gauged {{Gaussian}} Projected Entangled Pair States: {{A}} High Dimensional Tensor Network Formulation for Lattice Gauge Theories},
  shorttitle = {Gauged {{Gaussian}} Projected Entangled Pair States},
  author = {Kelman, Ariel and Borla, Umberto and Gomelski, Itay and Elyovich, Jonathan and Roose, Gertian and Emonts, Patrick and Zohar, Erez},
  year = 2024,
  month = sep,
  journal = {Physical Review D},
  volume = {110},
  number = {5},
  pages = {054511},
  publisher = {American Physical Society},
  doi = {10.1103/PhysRevD.110.054511},
  urldate = {2024-10-14}
}

@misc{Kelman-GGPEPSGithubSource-2026,
  title = {{{GGPEPS Github Source Code}}},
  author = {Kelman, Ariel and Emonts, Patrick and Gomelski, Itay and Elyovich, Jonathan},
  year = 2026,
  month = jul,
  urldate = {2026-08-26},
  url = {https://github.com/akelman/ggpeps},
  copyright = {MIT}
}

@misc{Kelman-GgpepsPackageGauged-2026,
  title = {Ggpeps: {{A Package}} for {{Gauged Gaussian PEPS}}},
  shorttitle = {Ggpeps},
  author = {Kelman, Ariel and Gomelski, Itay and Elyovich, Jonathan and Zohar, Erez and Emonts, Patrick},
  year = 2026,
  month = sep,
  number = {arXiv:2609.06486},
  eprint = {2609.06486},
  primaryclass = {hep-lat},
  publisher = {arXiv},
  doi = {10.48550/arXiv.2609.06486},
  urldate = {2026-09-09},
  archiveprefix = {arXiv}
}

@article{Kogut-HamiltonianFormulationWilsons-1975,
  title = {Hamiltonian Formulation of {{Wilson}}'s Lattice Gauge Theories},
  author = {Kogut, John and Susskind, Leonard},
  year = 1975,
  month = jan,
  journal = {Physical Review D},
  volume = {11},
  number = {2},
  pages = {395--408},
  publisher = {American Physical Society},
  doi = {10.1103/PhysRevD.11.395},
  urldate = {2022-11-27}
}

@article{kraus_fermionic_2010,
  title = {Fermionic Projected Entangled Pair States},
  author = {Kraus, Christina V. and Schuch, Norbert and Verstraete, Frank and Cirac, J. Ignacio},
  year = 2010,
  month = may,
  journal = {Physical Review A},
  volume = {81},
  number = {5},
  pages = {052338},
  issn = {1050-2947, 1094-1622},
  doi = {10.1103/PhysRevA.81.052338},
  urldate = {2022-11-14},
  langid = {english}
}

@book{Langacker-StandardModel-2017,
  title = {The {{Standard Model}} and {{Beyond}}},
  author = {Langacker, Paul},
  year = 2017,
  month = jul,
  edition = {2},
  publisher = {CRC Press}
}

@article{magnifico_lattice_2021,
  title = {Lattice Quantum Electrodynamics in (3+1)-Dimensions at Finite Density with Tensor Networks},
  author = {Magnifico, Giuseppe and Felser, Timo and Silvi, Pietro and Montangero, Simone},
  year = 2021,
  month = jun,
  journal = {Nature Communications},
  volume = {12},
  number = {1},
  pages = {3600},
  publisher = {Nature Publishing Group},
  issn = {2041-1723},
  doi = {10.1038/s41467-021-23646-3},
  urldate = {2022-09-09},
  copyright = {2021 The Author(s)},
  langid = {english}
}

@article{montangero_loop-free_2022,
  title = {Loop-Free Tensor Networks for High-Energy Physics},
  author = {Montangero, Simone and Rico, Enrique and Silvi, Pietro},
  year = 2022,
  month = feb,
  journal = {Philosophical Transactions of the Royal Society A: Mathematical, Physical and Engineering Sciences},
  volume = {380},
  number = {2216},
  pages = {20210065},
  issn = {1364-503X, 1471-2962},
  doi = {10.1098/rsta.2021.0065},
  urldate = {2022-02-28},
  langid = {english}
}

@article{orus_practical_2014,
  ids = {orus_practical_2014-1},
  title = {A Practical Introduction to Tensor Networks: {{Matrix}} Product States and Projected Entangled Pair States},
  shorttitle = {A Practical Introduction to Tensor Networks},
  author = {Or{\'u}s, Rom{\'a}n},
  year = 2014,
  month = oct,
  journal = {Annals of Physics},
  volume = {349},
  pages = {117--158},
  issn = {00034916},
  doi = {10.1016/j.aop.2014.06.013},
  urldate = {2017-07-04}
}

@incollection{Pannullo-LatticeInvestigationPhase-2020,
  title = {Lattice Investigation of the Phase Diagram of the 1+1 Dimensional {{Gross-Neveu}} Model at Finite Number of Fermion Flavors},
  booktitle = {Proceedings of 37th {{International Symposium}} on {{Lattice Field Theory}} --- {{PoS}}({{LATTICE2019}})},
  author = {Pannullo, Laurin and Lenz, Julian and Wagner, Marc and Wellegehausen, Bj{\"o}rn and Wipf, Andreas},
  year = 2020,
  month = aug,
  volume = {363},
  pages = {063},
  publisher = {SISSA Medialab},
  doi = {10.22323/1.363.0063},
  urldate = {2026-11-12},
  langid = {english}
}

@book{peskin_introduction_1995,
  ids = {peskin_introduction_1995-1},
  title = {An Introduction to Quantum Field Theory},
  author = {Peskin, Michael Edward and Schroeder, Daniel V.},
  year = 1995,
  publisher = {Addison-Wesley Pub. Co},
  address = {Reading, Mass},
  isbn = {978-0-201-50397-5},
  lccn = {QC174.45 .P465 1995}
}

@article{Polyakov-CompactGaugeFields-1975,
  title = {Compact Gauge Fields and the Infrared Catastrophe},
  author = {Polyakov, A. M.},
  year = 1975,
  month = oct,
  journal = {Physics Letters B},
  volume = {59},
  number = {1},
  pages = {82--84},
  issn = {0370-2693},
  doi = {10.1016/0370-2693(75)90162-8},
  urldate = {2026-11-04}
}

@article{Rico-TensorNetworksLattice-2014,
  title = {Tensor {{Networks}} for {{Lattice Gauge Theories}} and {{Atomic Quantum Simulation}}},
  author = {Rico, E. and Pichler, T. and Dalmonte, M. and Zoller, P. and Montangero, S.},
  year = 2014,
  month = may,
  journal = {Physical Review Letters},
  volume = {112},
  number = {20},
  pages = {201601},
  publisher = {American Physical Society},
  doi = {10.1103/PhysRevLett.112.201601},
  urldate = {2026-11-22}
}

@article{robaina_simulating_2021,
  title = {Simulating 2 + 1 {{D Z}} 3 {{Lattice Gauge Theory}} with an {{Infinite Projected Entangled-Pair State}}},
  author = {Robaina, Daniel and Ba{\~n}uls, Mari Carmen and Cirac, J. Ignacio},
  year = 2021,
  month = feb,
  journal = {Physical Review Letters},
  volume = {126},
  number = {5},
  pages = {050401},
  publisher = {American Physical Society (APS)},
  issn = {0031-9007, 1079-7114},
  doi = {10.1103/PhysRevLett.126.050401},
  urldate = {2022-11-14},
  langid = {english}
}

@article{Roose-GaugingSuperpositionFermionic-2025,
  title = {Gauging a Superposition of Fermionic {{Gaussian}} Projected Entangled Pair States to Get Lattice Gauge Theory Eigenstates},
  author = {Roose, Gertian and Zohar, Erez},
  year = 2025,
  month = nov,
  journal = {Journal of High Energy Physics},
  volume = {2025},
  number = {11},
  pages = {52},
  issn = {1029-8479},
  doi = {10.1007/JHEP11(2025)052},
  urldate = {2026-06-11},
  langid = {english}
}

@article{Roose-LatticeRegularisationEntanglement-2021,
  title = {Lattice Regularisation and Entanglement Structure of the {{Gross-Neveu}} Model},
  author = {Roose, Gertian and Bultinck, Nick and Vanderstraeten, Laurens and Verstraete, Frank and Van Acoleyen, Karel and Haegeman, Jutho},
  year = 2021,
  month = jul,
  journal = {Journal of High Energy Physics},
  volume = {2021},
  number = {7},
  pages = {207},
  issn = {1029-8479},
  doi = {10.1007/JHEP07(2021)207},
  urldate = {2026-11-12},
  langid = {english}
}

@misc{rouxinol_quantum_2026,
  title = {Quantum {{Phase Diagram}} of the \$2+1\${{D Untruncated SU}}\$(2)\$ {{Lattice Gauge Theory}} with {{Dynamical Fermions}}},
  author = {Rouxinol, Gabriel and Bender, Julian and Emonts, Patrick and Grossi, Michele and Halimeh, Jad C.},
  year = 2026,
  month = jul,
  number = {arXiv:2607.26132},
  eprint = {2607.26132},
  primaryclass = {hep-lat},
  publisher = {arXiv},
  doi = {10.48550/arXiv.2607.26132},
  urldate = {2026-08-09},
  archiveprefix = {arXiv}
}

@article{schollwock_density-matrix_2011,
  ids = {schollwock_density-matrix_2011-1},
  title = {The Density-Matrix Renormalization Group in the Age of Matrix Product States},
  author = {Schollw{\"o}ck, Ulrich},
  year = 2011,
  month = jan,
  journal = {Annals of Physics},
  volume = {326},
  number = {1},
  pages = {96--192},
  issn = {00034916},
  doi = {10.1016/j.aop.2010.09.012},
  urldate = {2017-06-22}
}

@article{sorella_generalized_2001,
  title = {Generalized {{Lanczos}} Algorithm for Variational Quantum {{Monte Carlo}}},
  author = {Sorella, Sandro},
  year = 2001,
  month = jun,
  journal = {Physical Review B},
  volume = {64},
  number = {2},
  pages = {024512},
  issn = {0163-1829, 1095-3795},
  doi = {10.1103/PhysRevB.64.024512},
  urldate = {2022-11-14},
  langid = {english}
}

@article{sorella_wave_2005,
  title = {Wave Function Optimization in the Variational {{Monte Carlo}} Method},
  author = {Sorella, Sandro},
  year = 2005,
  month = jun,
  journal = {Physical Review B},
  volume = {71},
  number = {24},
  pages = {241103},
  issn = {1098-0121, 1550-235X},
  doi = {10.1103/PhysRevB.71.241103},
  urldate = {2022-11-14},
  langid = {english}
}

@misc{Spriggs-AccurateGroundStates-2025,
  title = {Accurate Ground States of \${{SU}}(2)\$ Lattice Gauge Theory in 2+{{1D}} and 3+{{1D}}},
  author = {Spriggs, Thomas and Greplova, Eliska and Carrasquilla, Juan and Nys, Jannes},
  year = 2025,
  month = sep,
  number = {arXiv:2509.12323},
  eprint = {2509.12323},
  primaryclass = {hep-lat},
  publisher = {arXiv},
  doi = {10.48550/arXiv.2509.12323},
  urldate = {2026-11-22},
  archiveprefix = {arXiv}
}

@article{susskind_lattice_1977,
  ids = {susskind_lattice_1977-1},
  title = {Lattice Fermions},
  author = {Susskind, Leonard},
  year = 1977,
  month = nov,
  journal = {Physical Review D},
  volume = {16},
  number = {10},
  pages = {3031--3039},
  publisher = {American Physical Society},
  issn = {0556-2821},
  doi = {10.1103/PhysRevD.16.3031},
  urldate = {2017-12-04}
}

@article{tagliacozzo_entanglement_2011,
  title = {Entanglement Renormalization and Gauge Symmetry},
  author = {Tagliacozzo, L. and Vidal, G.},
  year = 2011,
  month = mar,
  journal = {Physical Review B},
  volume = {83},
  number = {11},
  pages = {115127},
  doi = {10.1103/PhysRevB.83.115127},
  urldate = {2019-12-02}
}

@article{Tagliacozzo-TensorNetworksLattice-2014,
  title = {Tensor {{Networks}} for {{Lattice Gauge Theories}} with {{Continuous Groups}}},
  author = {Tagliacozzo, L. and Celi, A. and Lewenstein, M.},
  year = 2014,
  month = nov,
  journal = {Physical Review X},
  volume = {4},
  number = {4},
  pages = {041024},
  publisher = {American Physical Society},
  doi = {10.1103/PhysRevX.4.041024},
  urldate = {2026-11-22}
}

@article{Troyer-ComputationalComplexityFundamental-2005,
  title = {Computational {{Complexity}} and {{Fundamental Limitations}} to {{Fermionic Quantum Monte Carlo Simulations}}},
  author = {Troyer, Matthias and Wiese, Uwe-Jens},
  year = 2005,
  month = may,
  journal = {Physical Review Letters},
  volume = {94},
  number = {17},
  pages = {170201},
  publisher = {American Physical Society},
  doi = {10.1103/PhysRevLett.94.170201},
  urldate = {2023-03-28}
}

@article{tschirsich_phase_2019,
  ids = {tschirsich_phase_2019-1,tschirsich_phase_2019-2},
  title = {Phase Diagram and Conformal String Excitations of Square Ice Using Gauge Invariant Matrix Product States},
  author = {Tschirsich, Ferdinand and Montangero, Simone and Dalmonte, Marcello},
  year = 2019,
  month = mar,
  journal = {SciPost Physics},
  volume = {6},
  number = {3},
  publisher = {SciPost},
  issn = {2542-4653},
  doi = {10.21468/SciPostPhys.6.3.028},
  urldate = {2019-06-26}
}

@article{verstraete_matrix_2004,
  title = {Matrix {{Product Density Operators}}: {{Simulation}} of {{Finite-Temperature}} and {{Dissipative Systems}}},
  shorttitle = {Matrix {{Product Density Operators}}},
  author = {Verstraete, F. and {Garc{\'i}a-Ripoll}, J. J. and Cirac, J. I.},
  year = 2004,
  month = nov,
  journal = {Physical Review Letters},
  volume = {93},
  number = {20},
  pages = {207204},
  issn = {0031-9007, 1079-7114},
  doi = {10.1103/PhysRevLett.93.207204},
  urldate = {2022-11-14},
  langid = {english}
}

@misc{Verstraete-RenormalizationAlgorithmsQuantumMany-2004,
  title = {Renormalization Algorithms for {{Quantum-Many Body Systems}} in Two and Higher Dimensions},
  author = {Verstraete, F. and Cirac, J. I.},
  year = 2004,
  month = jul,
  number = {arXiv:cond-mat/0407066},
  eprint = {cond-mat/0407066},
  publisher = {arXiv},
  doi = {10.48550/arXiv.cond-mat/0407066},
  urldate = {2023-04-03},
  archiveprefix = {arXiv}
}

@article{White-DensitymatrixAlgorithmsQuantum-1993,
  title = {Density-Matrix Algorithms for Quantum Renormalization Groups},
  author = {White, Steven R.},
  year = 1993,
  month = oct,
  journal = {Physical Review B},
  volume = {48},
  number = {14},
  pages = {10345--10356},
  publisher = {American Physical Society},
  doi = {10.1103/PhysRevB.48.10345},
  urldate = {2026-06-09}
}

@article{White-DensityMatrixFormulation-1992,
  title = {Density Matrix Formulation for Quantum Renormalization Groups},
  author = {White, Steven R.},
  year = 1992,
  month = nov,
  journal = {Physical Review Letters},
  volume = {69},
  number = {19},
  pages = {2863--2866},
  publisher = {American Physical Society},
  doi = {10.1103/PhysRevLett.69.2863},
  urldate = {2023-03-23}
}

@misc{zenodo_23889660,
  title = {Data for "Sidestepping the Sign Problem with Tensor Networks for Lattice Gauge Theory"},
  author = {Kelman, Ariel and Roose, Gertian and Emonts, Patrick and Zohar, Roose},
  year = 2026,
  month = nov,
  publisher = {Zenodo},
  doi = {10.5281/zenodo.23889660}
}

@article{zohar_building_2016,
  ids = {zohar_building_2016-1,zohar_building_2016-2},
  title = {Building Projected Entangled Pair States with a Local Gauge Symmetry},
  author = {Zohar, Erez and Burrello, Michele},
  year = 2016,
  month = apr,
  journal = {New Journal of Physics},
  volume = {18},
  number = {4},
  pages = {043008},
  publisher = {IOP Publishing},
  issn = {1367-2630},
  doi = {10.1088/1367-2630/18/4/043008},
  urldate = {2019-06-07}
}

@article{zohar_fermionic_2015,
  ids = {zohar_fermionic_2015-1},
  title = {Fermionic Projected Entangled Pair States and Local {{U}}(1) Gauge Theories},
  author = {Zohar, Erez and Burrello, Michele and Wahl, Thorsten B. and Cirac, J. Ignacio},
  year = 2015,
  month = dec,
  journal = {Annals of Physics},
  volume = {363},
  pages = {385--439},
  publisher = {Elsevier BV},
  issn = {00034916},
  doi = {10.1016/j.aop.2015.10.009},
  urldate = {2017-11-17}
}

@article{zohar_projected_2016,
  ids = {zohar_projected_2016-1,zohar_projected_2016-2},
  title = {Projected {{Entangled Pair States}} with Non-{{Abelian}} Gauge Symmetries: {{An SU}}(2) Study},
  shorttitle = {Projected {{Entangled Pair States}} with Non-{{Abelian}} Gauge Symmetries},
  author = {Zohar, Erez and Wahl, Thorsten B. and Burrello, Michele and Cirac, J. Ignacio},
  year = 2016,
  month = nov,
  journal = {Annals of Physics},
  volume = {374},
  pages = {84--137},
  publisher = {Elsevier BV},
  issn = {00034916},
  doi = {10.1016/j.aop.2016.08.008},
  urldate = {2018-01-24}
}

\end{document}